\documentstyle[twoside,fleqn,epsf,espcrc2]{article}
 
\def\be{\begin{equation}}
\def\ee{\end{equation}}
\def\bear{\begin{eqnarray}}
\def\eear{\end{eqnarray}}
\newcommand {\ber} {\begin{eqnarray*}}
\newcommand {\eer} {\end{eqnarray*}}

\newcommand {\eqref} [1] {(\ref {#1})}

\def\l{{\lambda}}

\def\t{{\tau}}

\def\ce{\centerline } 
\def\lra{$\Leftrightarrow$}

\def\th{theory} \def\Ft{F theory\ }

\def \B{type\ $II_B$\ }
\def \A{type\ $II_A$\ }
\def\bZ{{\bf Z}}

\def\bT{{\bf T}}
\def\M{M-theory\ }

\def\ra{\rightarrow}
\def\FL{(-1)^{F_L}}
\def \SL{$SL(2,{\cal Z})$}

\def\boxit#1{\vbox{\hrule\hbox{\vrule\kern3pt
\vbox{\kern3pt#1\kern3pt}\kern3pt\vrule}\hrule}}

\def\bT{{\bf T}}                                

\def\ce{\centerline } 
\def\lra{$\Leftrightarrow$}

\def\bCP{{\bf CP}} 

\def\ha{{1\over 2}}  
\def\cmp#1{{\it Comm. Math. Phys.} {\bf #1}}

\def\pl#1{{\it Phys. Lett.} {\bf #1B}}
\def\prl#1{{\it Phys. Rev. Lett.} {\bf #1}}

\def\np#1{{\it Nucl. Phys.} {\bf B#1}}

\def\jmath#1{{\it J. Math. Phys.} {\bf #1}}

\title{ Novel field theory phenomena from  
F theory and D-branes }
\author{ J. Sonnenschein 
\address{ School of Physics and Astronomy,
 Beverly and Raymond-Sackler Faculty of Exact Sciences,
 Tel-Aviv University, Ramat-Aviv, Tel-Aviv 69978, Israel}
\thanks{Work supported
    in part by the US-Israel Binational Science Foundation, by GIF --
    the German-Israeli Foundation for Scientific Research, and by the
    Israel Science Foundation.}}

\begin{document}
\begin{abstract}
Talk presented in the 97 Karpatcz winter school.
We describe Sen's work on F theory on K3 and its reflection
on the world-volume field theory on a D3-brane probe.
Field theories on a multiple of probes are analyzed. 
We construct  a 4d N=1 superconformal  probe theory which
is invariant under electric-magnetic duality via a   
 compactification 
to six dimensions on  $\bT^6/\bZ_2\times \bZ_2$. 
\end{abstract}
\maketitle
\section {Introduction}

Recent developments in string theories, and in particular brane
physcis, have provided  new magnifying glasses for observations 
on quantum field theories in various dimensions.
In this series of talks presented in the Karpatcz 97 winter school
we have focused on exploring field theories on the world-volume of
D-brane probes  in F theory\cite{vafa}. We have
followed the analysis
of A. Sen\cite{sen} on F theory on K3 and its reinterpretation
in terms of a probe theory\cite{bds}. 

 General background on D-branes, \B
theory, orientifolds and F theory is presented in the introduction
section. Section 2 is devoted to the analysis  of \Ft
compactified on K3\cite{sen} and its relation to 
the field theory on a D3-brane probe\cite{BDS}. 
A generalization of the latter to the theory on a multiple probe
system that is put in various configurations with respect to the
orientifold plane\cite{ASTY} is presented in section 3. 
Section 4 is devoted to the extraction of  the properties of 
a probe theory in the neighbourhood of an intersecting orietifold
planes of an F compactified on  $\bT^6/\bZ_2\times \bZ_2$. 
The field content is found using a Gimon-Polchinski\cite{gp} type
of analysis and a superpotential of this $N=1$ theory, which
admits a fixed line passing through the origin of the space of
couplings, is written down.  
The last 2 sections are 
 based on a work that was done in collaboration with
O. Aharony, S. Theisen and S. Yankielowicz\cite{ASTY}, 
 
\subsection{ A brief review of D-branes}
A Dp-brane is a p dimensional extended object prescribed by the
property that ends of open strings can lie on it, although  open
strings cannot exist in the bulk.\cite{pol} 
These open strings have Neuman boundary conditions (b.c)
$\partial_n X^\mu=0$ for $\mu=0,1,...,p$ and Dirichlet b.c
 $X^\mu=constant$ for $\mu=p+1,...9$.   

 We summarize several properties of D-branes that we will
make use of \cite{pcj}: 
(i) The D-brane is a dynamical object. It can fluctuate in
transverse directions.
Its world-volume theory includes gauge fields and scalar fields.
The latter correspond to the fluctuations.
 (ii) A Dp-brane couples to a $A_{p+1}$ form.   Thus type $II_A$
includes even branes and $II_B$ odd ones.
(iii) A T-duality on $S^1$ interchanges the Dirichlet and Neuman 
b.c on the $S^1$, so that a wrapped Dp-brane $\rightarrow$
D(p-1)-brane and unwrapped Dp-brane $\rightarrow$ D(p+1)-brane.
(iv) a D-brane is a BPS state. It breaks half of the
supersymmetries. (v) There is no force between infinite
parallel  D-branes  at rest. (vi) N parallel D-branes can form 
a bound state at threshold.

\subsection{ A. brief review of \B string theory}
We  briefly summarize the  bosonic fields and the symmetries of 
\B string theory.
\begin{enumerate}
\item {\bf Massless bosonic fields:} 

Neveu-Schwarz (NS-NS) sector- 
 $\phi$,\  $g_{\mu\nu}$,\ $B^{(NS)}_{\mu\nu}$.  

Ramond-Ramond (RR) sector- 
 $a$, \ $B^{(RR)}_{\mu\nu}$, $D^+_{\mu\nu\rho\sigma}$

\item {\bf Symmetries}

I. Symmetries of the perturbative \th-

(i) {\bf  $(-1)^{F_L}$} 

$a\ra -a$; $B^{(RR)}_{\mu\nu}\ra -B^{(RR)}_{\mu\nu}$; 
$D^+_{\mu\nu\rho\sigma}\ra -D^+_{\mu\nu\rho\sigma}$

(ii) {\bf  $\Omega$ world sheet parity transformation}-

$a\ra -a$;  $B^{(NS)}_{\mu\nu}\ra -B^{(NS)}_{\mu\nu}$; 
$D^+_{\mu\nu\rho\sigma}\ra -D^+_{\mu\nu\rho\sigma}$

II. conjectured non-perturbative  symmetry

\SL  S-duality-

$$
\fbox{$\displaystyle\l\ra {p\l+q\over r\l+s}\ \   
\left(\matrix{{B^{(NS)}_{\mu\nu}\ }\cr {B^{(RR)}_{\mu\nu}\ }\cr}\right)
\ra \left(\matrix{{p\ }{q\ }\cr
{r\ }{s\ }\cr}\right)
\left(\matrix{{B^{(NS)}_{\mu\nu} }\cr 
{B^{(RR)}_{\mu\nu}\ }\cr}\right) $}
$$

where 
$$\l =a+ie^{-\phi}$$

$p,q,r,s\in {\cal Z}$ with $ps-qr =1.$

\end{enumerate}
\subsection{ A brief review of \B orientifolds }

Type $II_B$  defined on ${{\cal M}\over G}$ 
where ${\cal M}$-manifold and  $G$- a group containing at least one element
of the form 

 $$g_{s-t}\cdot g_{int}\cdot \Omega$$
where $g_{s-t}$ is a space-time transformation,
$g_{int}$- is an  internal transformation and  $\Omega$
is a world-sheet parity  transformation. 

Here are some examples of \B orientiflods
\begin{enumerate}
\item \B on ${M^{9,1}\over \{1,\Omega \} }$
 This is type I string theory on $M^{9,1}$.
 
 \item \B on $M^{7,1}\times{T^2\over {\cal I}_{89}
(-1)^{F_L} \Omega }$
$ {\cal I}_{89}$ transforms $x^8\ra -x^8\ \ \ x^9\ra -x^9$

Upon performing T-duality on $T^2$ this is type I on  
$M^{7,1}\times T^2$.

\item \B on $M^{5,1}\times {T^4\over ({\cal I}_{67}
(-1)^{F_L} \Omega)\times {\cal I}_{89}
(-1)^{F_L} \Omega)}$

\end{enumerate}
\subsection{ A brief review of \Ft }

In  conventional compactifications of \B   
$\l=constant$.

\Ft is defined in the following way.
Let  ${\cal M}$ be a manifold of dimension $d+2$ 
admitting an elliptic fibration.
Let  ${\cal B}$ be a  (base) manifold of dimension d. 
${\cal M}$ is obtained by erecting at every point of 
${\cal B}$ a copy of a torus $T^2$ with its complex structure 
varying over ${\cal B}$.
\begin{figure}[h]
\hbox to\hsize{\hss
\epsfysize=3cm
\epsffile{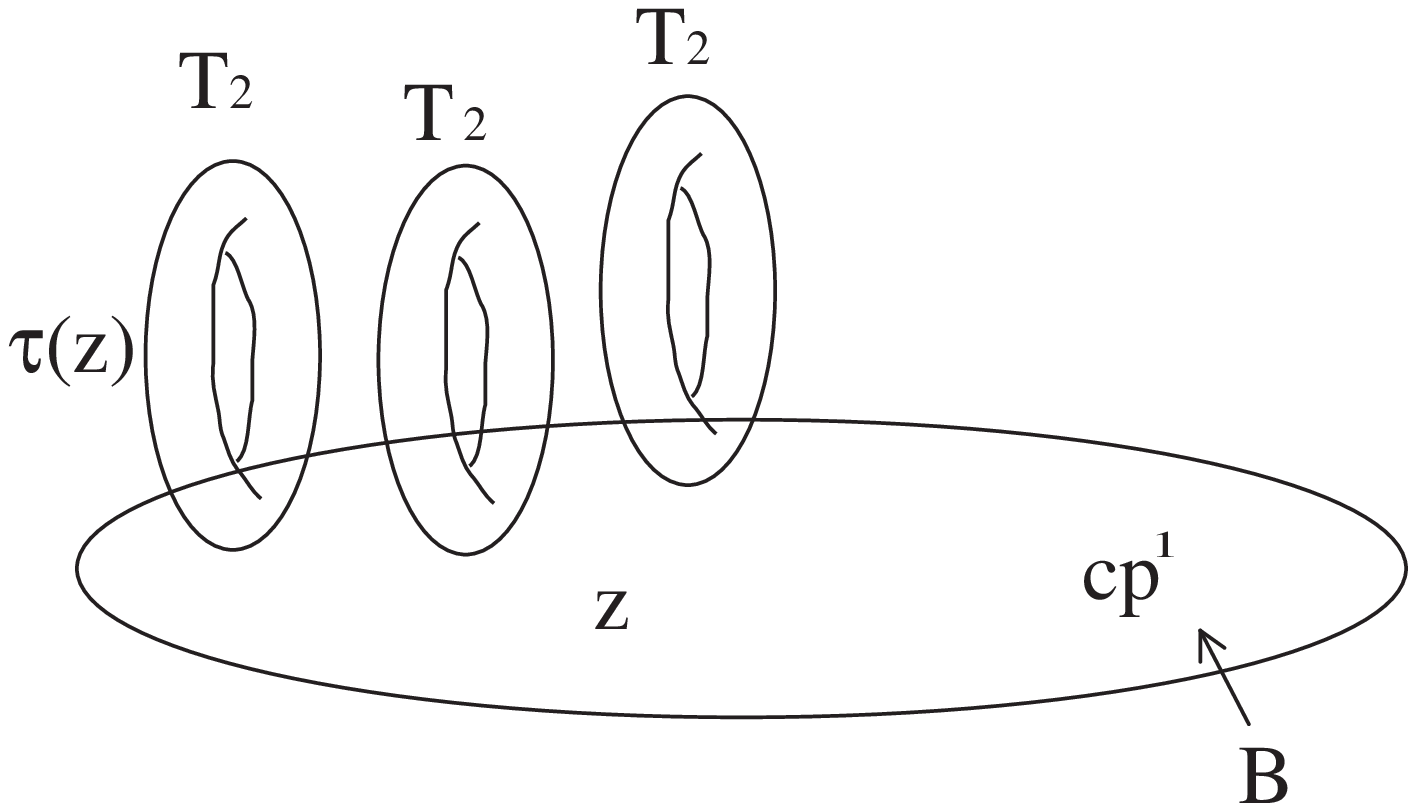} \hss}
\vspace{0cm}
\caption{} 
\end{figure}

$$
\fbox{$\displaystyle$
 \Ft on ${\cal M}$\  $\equiv$ \B on ${\cal B}$ }
$$
with $$\l=\tau(z)\neq\ constant$$

Note that when one moves along a closed cycle  $\l=\tau(z)$ 
must come to its orignal value up to an \SL transformation.
An example of elliptic fibration- elliptically fibered $K3$.
In this case ${\cal M}$- $K3$ and  ${\cal B}$-$CP^1$.
For $z$  the coordinate of $CP^1$
the torus is parametrized by the complex coordinates $x$ and $y$
\be
\label{mishyz} y^2= x^3+f_8(z)x+g_{12}(z)
\ee
$f_8(z)$ and $g_{12}(z)$ are polynomials of degree 8 and 12
respectively.

The $J$ invariant of $T^2$  is given by
\be
\label{mishj}j(\tau(z))={4(24f)^3\over 27g^2+ 4f^3}
\ee
\Ft on $K3$ is equivalent to \B on $CP^1$ with $\l=\t(z)$.
We enlist here several 
{\bf Dualities of \Ft} that will be useful in the discussions below.
\begin{enumerate}
\item  \Ft on ${T^2\over Z_2}$\  \lra\  type I (heterotic) $SO(32)$ 

\item \Ft on ${\cal M}\times S_1$\lra \B on ${\cal B}\times S_1$
with $\l(z) =\t(z)$ of the fibers.
Recall that \B on $ S_1$ with radius $R$ is dual to \A on  on $ S_1$ with
radius ${1\over R}$. Furthermore, 
\A on $ S_1$ is equivalent to \M on $T^2$,
so that $\l$ of the \B theory is identified with  $\t$ of the latter
$T^2$.  We thus conclude that 
$$\fbox{$\displaystyle$\Ft on ${\cal M}\times S_1$\lra \M 
on ${\cal M}$ }$$

\item  From (1) \Ft on $K3\times S_1$\lra \M on $K3$\lra Heterotic on
$T^3$.

In a similar manner, \Ft on $K3$\lra Heterotic on $T^2$.

Since the  heterotic theory  on $T^2$\lra type $I$ on  $T^2$\lra
\B on   ${T^2\over {\cal I}_{89}}$ one finds
$$\fbox{$\displaystyle$\Ft on $K3$\lra \B on  
${T^2\over {\cal I}_{89}}$}$$
\item \Ft on K3 fibered CY \lra heterotic on K3.

\end{enumerate}
\section{ Sen's model- F-theory on K3- space-time and 
worldvolume theories}
Eliptic fibration of K3  is described by eqn. \eqref{mishyz}.
The idea of Sen was to look for a point in the F-theory moduli space where 

{\bf $\tau(z)$ is independent of $z$ and $Im(\tau)$ is large}

This should imply a
conventional perturbative         
string compactification, and as will be shown below a 
 conformal invariant field theory on a ``probe" worldvolume.   

A constant $\tau(z)$ means also 
\be 
j(\tau(z))=constant\ra {f^3\over g^2}= constant
\ee

This condition is obeyed if one takes 
\be
\label{phi} f=\alpha\phi(z)^2,\ \ \  g= \phi(z)^3
\ee
In that case 
$j(\tau(z))= {4(24 \alpha)^3\over 27 + 4\alpha^3}=constant$ 
where 
$\alpha$ can be adjusted so that $Im(\tau)$ is large.

{\it Singularities} of $j(\tau)$ at 
\be
\Delta = (27g^2 + 4f^3)=0
\ee

For the ansatz \eqref{phi} and with $\phi \sim \prod_{i=1}^4(z-z_i)$
\be
\Delta \sim
 \prod_{i=1}^4(z-z_i)^6
\ee

{\it Metric on the base $CP^1$}

In general the metric is given by\cite{gsvy}
$$ ds^2 = F(\tau,\bar\tau))\prod_{i=1}^{24}(z-z_i)^{-{1\over 12}}
\prod_{i=1}^{24}(\bar z-\bar z_i)^{-{1\over 12}}$$

So for the point in moduli space we consider it takes the form  
$$\fbox{$\displaystyle ds^2 = F(\tau,\bar\tau))\prod_{i=1}^{4}(z-z_i)^{-{1\over 2}}
\prod_{i=1}^{4}(\bar z-\bar z_i)^{-{1\over 2}}$}$$

This can be brought to the form of a {\bf Flat metric}. Define 
$$dw=\prod_{i=1}^{4}(z-z_i)^{-{1\over 2}}dz$$

then 
$$ ds^2 = dwd\bar w. $$

Near $z=z_i$,\ \   $w\sim (z-z_i)^{-{1\over 2}}$, namely, there is 
a{\bf Conical deficit angle of $\pi$} at each of the four $z_i$.

 Naively, we may conclude that  

\be
{\cal B}= {T^2\over {\cal I}_{89}}
\ee

But there is a subtlety, lets check again 

\be
\fbox{$\displaystyle 
y^2= x^3 +\alpha x\prod_{i=1}^{4}(z-z_i)^2
+\prod_{i=1}^{4}(z-z_i)^3$}
\ee

going around an orbifold point like $z=z_i$ 

\ber
(z-z_i)\ra& e^{2\pi i}(z-z_i)\cr
x \ra& e^{2\pi i}x=x\cr
y \ra& e^{3\pi i}y=-y\cr
\eer
which is

$$\fbox{$\displaystyle An\ \SL\ trans.
 \pmatrix{-1&\cr &-1\cr}\ on\ the\ T^2
fiber$}$$

The transformation $\pmatrix{-1&\cr &-1\cr}$  corresponds to

$$\FL\Omega$$

Thus, the the base is infact  an {\bf Orientifold}

 $${\cal B}= {T^2\over \FL {\cal I}_{89}\Omega}.$$

Under this projection $z\ra -z$, 
so define $u=z^2$
which is single valued on the orientifold.

{\bf The orientifold} 

(1) preserves half of the space-time supersymmetries of \B .

(2) It carries a RR topological charge  $-4$\cite{pol}

\begin{figure}[h]
\hbox to\hsize{\hss
\epsfysize=3cm
\epsffile{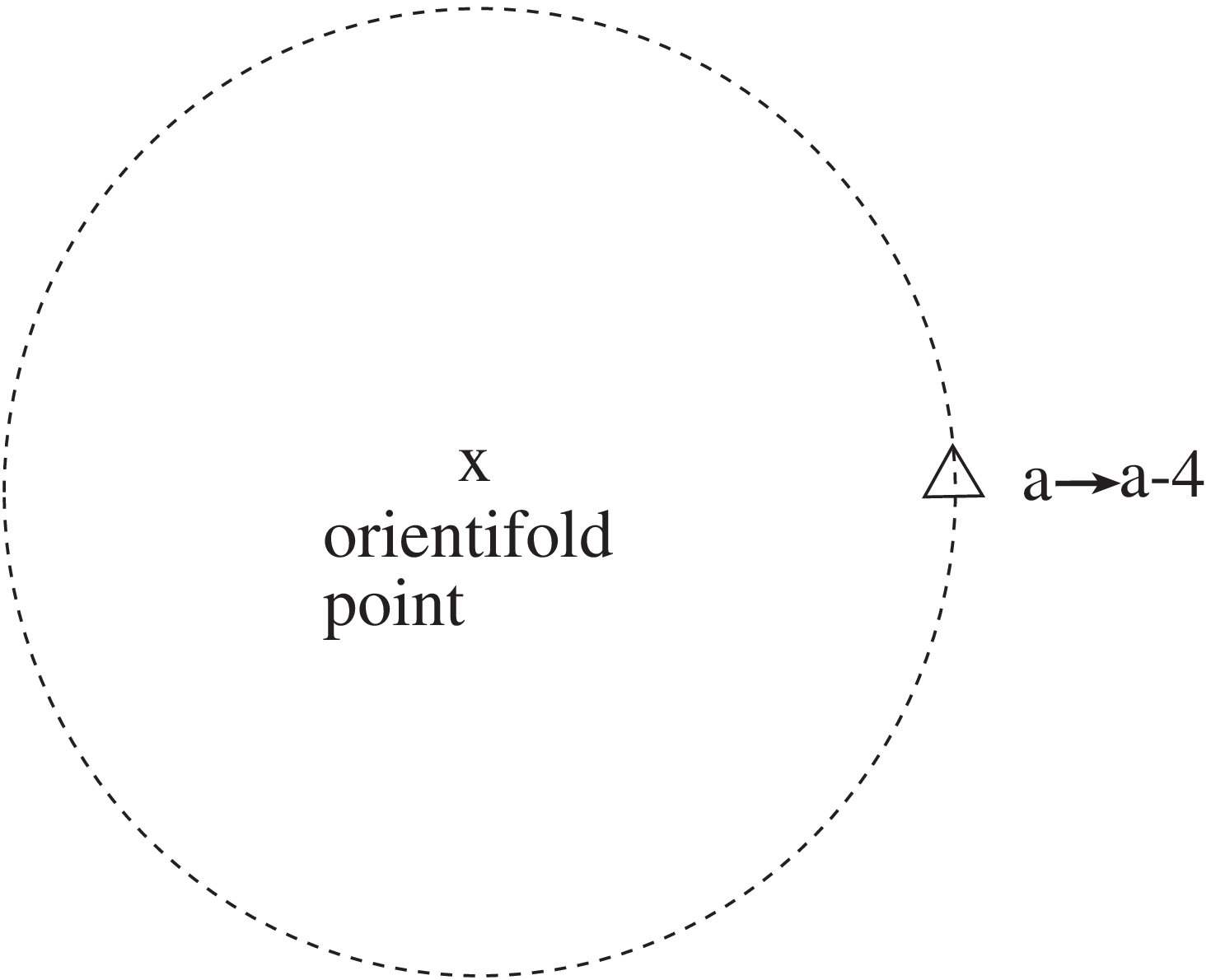} \hss}
\vspace{0cm}
\caption{} \label{pic1a}
\end{figure}

\be
\oint_C da =-4
\ee

where $C$ is a contour around $u=0$


{\bf D7-brane } 

(1) preserves half of the space-time supersymmetries of \B .

(2) It carries a RR topological charge  $+1$

\begin{figure}[h]
\hbox to\hsize{\hss
\epsfysize=3cm
\epsffile{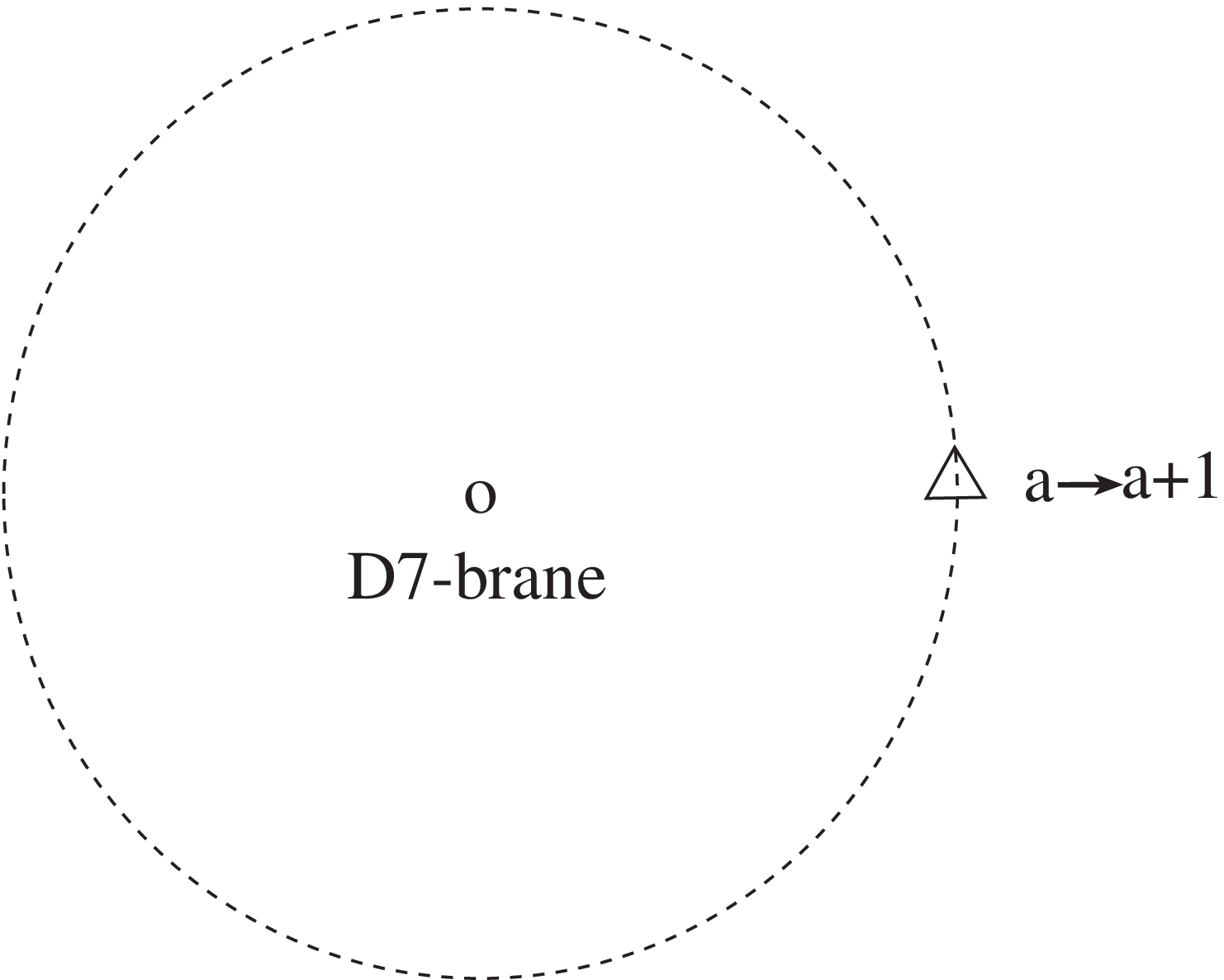} \hss}
\vspace{0cm}
\caption{} \label{pic1a}
\end{figure}

To cancell the RR charges of the 4 orientifolds the theory must have 
{\bf 16 
D7-branes} see fig. 4.


Since we have a configuration with 

$\lambda=constant\ra a=constant$ 
the RR charge must be cancelled {\bf locally} on the orientifold  
(fig. 5)
\begin{figure}[h]
\hbox to\hsize{\hss
\epsfysize=3cm
\epsffile{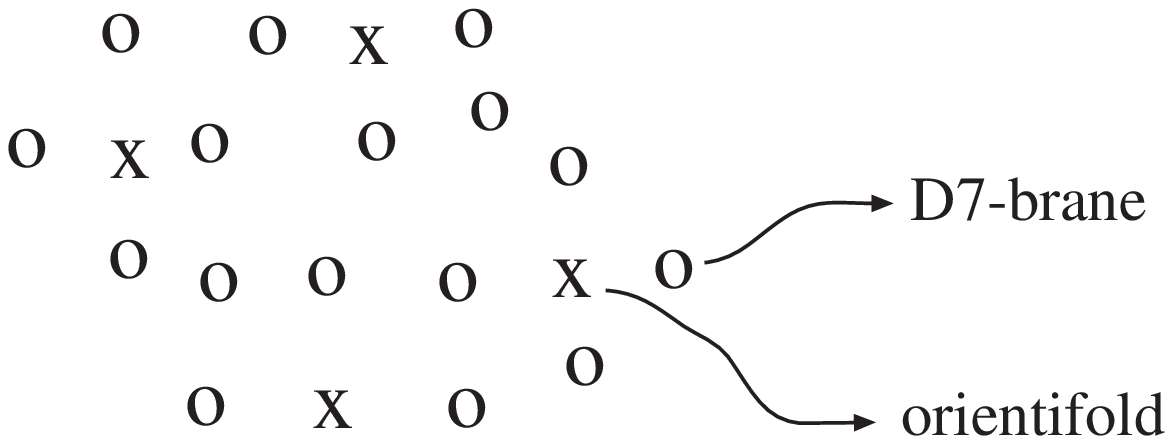} \hss}
\vspace{0cm}
\caption{} \label{pic1a}
\end{figure}

There is an {\bf Enhanced symmetry at singularies }
that can be viewed in
F-theory picture and in the  orientifold picture.

Start with F-theory picture

\be
y^2= x^3 +\alpha x\prod_{i=1}^{4}(z-z_i)^2
+\prod_{i=1}^{4}(z-z_i)^3
\ee

Near $z=z_1$

\be\tilde y^2= \tilde x^3 +\alpha \tilde x(z-z_1)^2
+(z-z_1)^3
\ee

which means that there is a 
$$\fbox{$\displaystyle$ $D_4$ type singularity\ $\ra\  SO(8)^4$
gauge symmetry }$$

Now the orientifold description
Let us construct the string configuration in several steps:
(i) From strings starting and ending on the 7Dbrane fig. 6. 
one finds
\begin{figure}[h]
\hbox to\hsize{\hss
\epsfysize=3cm
\epsffile{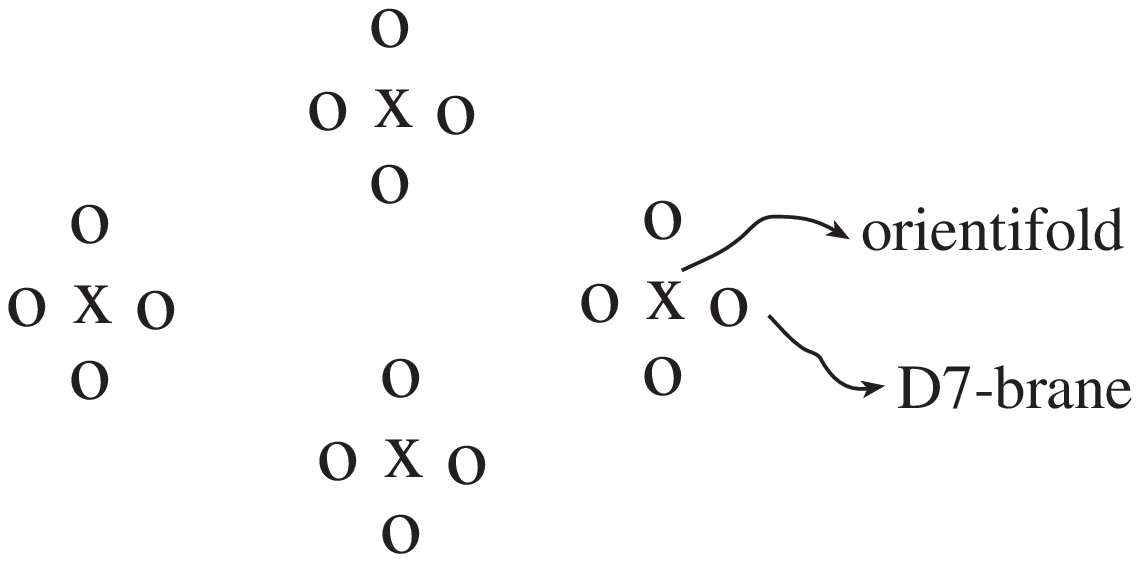} \hss}
\vspace{0cm}
\caption{}  \label{pic1a}
\end{figure}

 
$N=1$ 8D wv theory with $U(1)$ gauge field and a complex scalar 
that corresponds to the location of the D-brane in the $z$     
plane.

(ii) When two D-branes collapse  one gets (fig. 7)

\begin{figure}[h]
\hbox to\hsize{\hss
\epsfysize=3cm
\epsffile{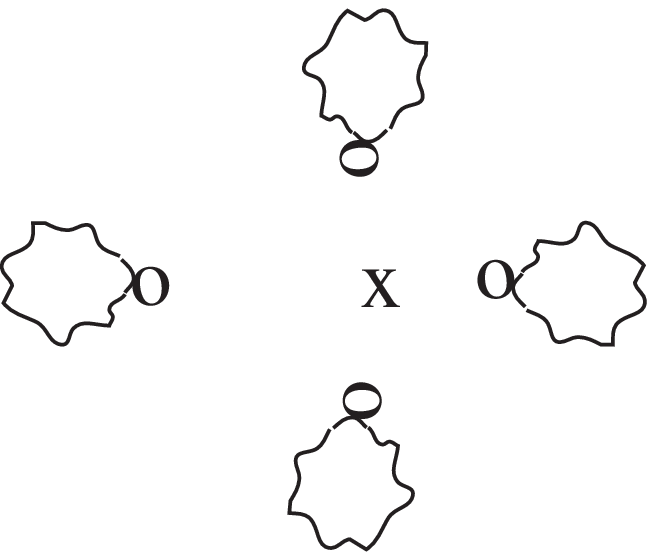} \hss}
\vspace{0cm}
\caption{} \label{pic1a}
\end{figure}


$N=1$ 8D wv theory with $U(2)$ gauge fields and 
a complex scalar in the adjoint.

(iii) When all the four $z_i$ are equal (fig. 8) then the 

\begin{figure}[h]
\hbox to\hsize{\hss
\epsfysize=3cm
\epsffile{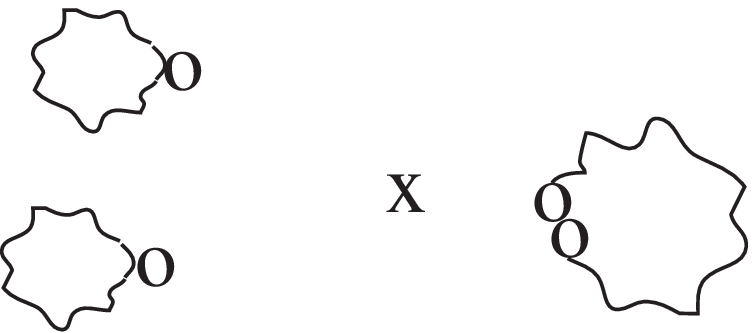} \hss}
\vspace{0cm}
\caption{} \label{pic1a}
\end{figure}
 symmetry is enhanced to $U(4)$

(iv) when all the four $z_i=0$ (fig. 9)

\begin{figure}[h]
\hbox to\hsize{\hss
\epsfysize=3cm
\epsffile{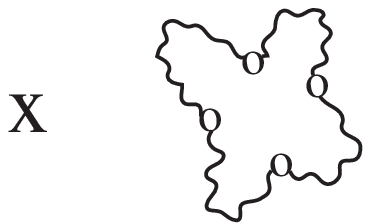} \hss}
\vspace{0cm}
\caption{} \label{pic1a}
\end{figure}

there are  
additional strings (and hence massless vectors) between the 7branes 
at $z_i$ and  their mirrors at $-z_i$.
Now the symmetry is enhanced to $U(8)$

(v) But the generator of $Z_2= \FL {\cal I}_{89}\Omega$
 acts on the generators $T^a$ of the $U(8)$. 

Only those that commute with it
survive. Under $Z_2$
$$\fbox{$\displaystyle T^a\ra_{Z_2} -\pmatrix{0\  I_4\ \cr I_4\ 0\ \cr}\  (T^a)^T\ 
 \pmatrix{0\  I_4\ \cr I_4\ 0\ \cr}$}$$
the various factors of this transformations are 
(i)-sign: change the sign of the oscillator $A^\mu_{-1}$ that
creates  open strings from the vaccum.

(ii)Tranaspose: effect of orientation reversal

(iii)$\pmatrix{0 I_4\cr I_4 0\cr}$ : effect of exchange of Dbranes
and  their images.  This breaks 
$$U(8)\ra SO(8)$$
Let us now move away from the special point in the  moduli space by
switching background fields in both the orientifold and F-theory
descriptions.

{\bf Deforming away from the special point}-
       
{\it Orientifold description}-
At each of the orientifolds the 7D-branes can be moved away from the
orientfold.
Near one particular orientifold the theory is \B on 
${{\cal R}^2\over \FL {\cal I}_{89}\Omega}$ and the picture is 
like that of the neighborhood of a D7-brane in fig. 5. 
\begin{figure}[h]
\hbox to\hsize{\hss
\epsfysize=3cm
\epsffile{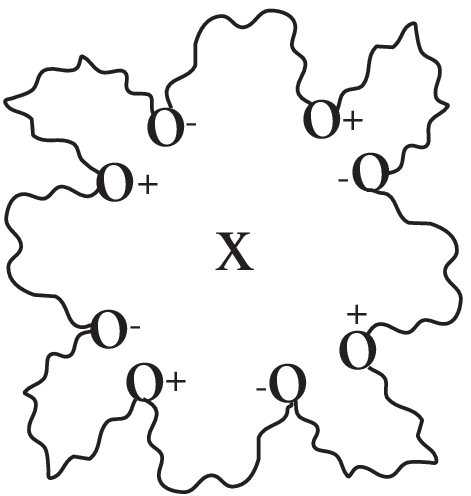} \hss}
\vspace{0cm}
\caption{} \label{pic1a}
\end{figure}

The background field $\lambda$

$$\fbox{$\displaystyle \l(z)=\l_0 + {1\over 2\pi i}
(\sum_{i=1}^4 \ln(z-z_i)- 4\ln z)$}$$
It is clear that 
 {\bf close to  $z=0$ solution not consistent} since $Im(\l)<0$.

{\it F-theory description}

Around the orbifold point
$$y^2= x^3 +f(z)x +g(z)$$
where $f(z)$ and $g(z)$ are polynomial of 
 degree 2  and  degree 3 respectively.

The  zeros of $\Delta$ are now at 6 singular points

\begin{figure}[h]
\hbox to\hsize{\hss
\epsfysize=3cm
\epsffile{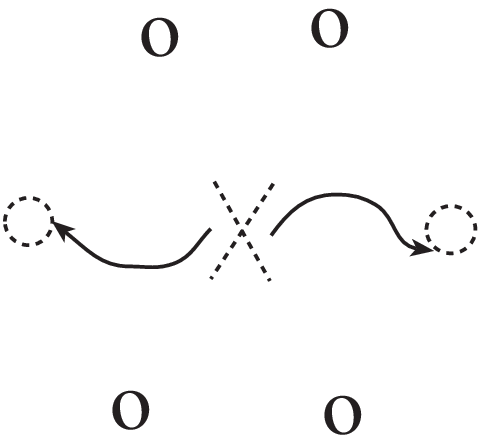} \hss}
\vspace{0cm}
\caption{} \label{pic1a}
\end{figure}

The total number of parameters is 
\ber
 &3\  from\  (f) +4\  from\  (g)\\
&-1
from\ rescaling\ of\ x, y\ -1 shift\ of\ z=5\\
\eer
Recall the Seiberg-Witten solution for $N=2$ $SU(2)$  
gauge symemtry
with $N_f=4$  is parametrized by
$$ \tau(z,\ \  m_1,\ \  m_2,\ \    m_3,\ \  m_4,\ \ \tau_0)$$
 where $z=<\phi^2>$, $m_i$ are  quark masses
and $\tau_0$  is the classical coupling. 
It is now natural to 
use the  SW parameters $z,\tau_0,m_i$ to label the F-theory
background.

In this description  the singularities are associated with 
massless quarks, massless monopole 
and massless dyon.  
\begin{center}
\begin{tabular}{|c|c|c|}\hline
 & orientifold                 & F-theory     \\ \hline   
 & & \\
para- &$\l_0,$  &$\t_0,$ \\
meters&$ z_1,z_2,z_3,z_4 $& $ m_1,m_2,m_3,m_4$\\
 & & \\ \hline
 & & \\
back-  & $\l(z)=$&$\l(z)=$\\
&$\l_0 +{1\over 2\pi i}(\sum_{i=1}^4$
& $ \t(\t_0, m_i)$\\
ground $\l$ & $ \ln(z-z_i)- 4\ln z)$ & \\ \hline 
 & & \\
asymp- & $\l_0$& $\t_0$\\ 
totic $\l$ & & \\  \hline
 & & \\
BPS  & $z_i$ & $m_i^2$\\ 
spectrum & & \\  \hline  \end{tabular} 
\end{center}

The orientifold solution for $\l(z)$
is the same as the SW solution for large $z$ ($u$)
$$\t(z)=\t_0 + {1\over 2\pi i}(\sum_{i=1}^4 \ln(z-m_i^2)- 4\ln z) $$
The singularities $\ln(z-m_i^2)$
are due to 
 the masselss quarks  and   the $4\ln z$  term 
is associated with  massless $SU(2)$
gauge bosons
which splits into two singularities in the full solution.
The comparison between the orientifold and the F theory picture  is
summarized in the table above.
 
We can now conclude that 
\boxit{ F-theory provides the quantum corrected version of the 
orientifold background in the form of {\bf Seiberg-Witten} solution}  

{\it Question}:

 Is it a conincidence that the SW theory emmerged?


\subsection{ SW solution as the wv theory on a 3B probe}

Let us study the effective Lagrangian on a probe that moves 
in the background of the F-theory on K3. 

{\it Question:} What probe should one use?
3Brane is the most natural probe since:
(1) 3Brane is invariant under \SL, so the wv theory should also be 
invariant under \SL duality,
(2) 7Brane parallel to a 3Brane probe breaks $1/2$ of the
supersymmetries so that   $\ra$ $N=2$ in 4D.

Another interpretation of the 3Brane probe is as a
{\bf wrapped 5Brane}.
Recall  F-theory on K3 is dual to type I on $T^2$.
Take a 5Brane of type I. Perform a T-duality on the $T^2$
which interchanges $N$\lra$ D$ boundary conditions, and you find 
 a 3D-Brane of type I'.
The Chan-Paton $SU(2)$ gauge group of the 5Brane of type I is
inherited  by the  3Brane of the type I'.
The comparison between the two dual pictures is summarized in the
following table\cite{bds}.
  \begin{center}
\begin{tabular}{|c|c|}\hline
 type I 5Brane                 & type I' 3Brane   \\ \hline   
  & \\
Wilson lines & 3Brane position \\
around cycles & on ${T^2\over Z_2}$ away \\
of $T^2$  &  from orientifold \\
$SU(2)\ra U(1)$&$SU(2)\ra U(1)$ \\ \hline

position of 7Brane & position of 7Brane\\

$R'_8\alpha_{i,8}+iR'_9\alpha_{i,9}$& $m_i$\\
position of 3Brane & position of 3Brane\\
$R'_8 A_8+iR'_9 A_9=\pmatrix{w& 0\cr 0&-w}$& $z=w^2$\\
  & \\  \hline  \end{tabular} 
\end{center}
where the notation in this table is that of ref. \cite{bds}.  
The strings streched from the 3Brane to the 7Brane
induce fundamental matter hypers with  mass $m_i\pm w$. 

Let us summarize 
 the correspondence between the \Ft and the probe
theory:

(i) The set-up on the 3Brane wv is that of $SU(2)$ SW theory with 
four hyper-multiplets in the fundamental rep.

(ii) The F-theory $\l$  $\ra$\  probe  gauge coupling $\t$.

(iii) The fact that   $\l=constant$\ \  $\ra$\ \  into confomal invarinace of 
the probe theory.

(iv) \B \SL invarinace \ $\ra$\ \SL invariace of $SU(2)$ with $N_f=4$   
 
\section{The world-volume field theory on a multiple of 3-branes  probes in 8
dimensions}

Consider again \Ft on K3.
Recall that \Ft on K3 is dual to type $I$ on $T^2$.
Now wrap $N_c$ 5B around the $T^2$. Let us analyze the properties 
of a world-volume theory on the muptiple probes \cite{dls}.

(i){\it Supersymmetry}- 
8 susy charges $N=1$ in 5+1 dim$\ra$ $N=2$  in 3B 3+1 wv.

(ii){\it Gauge symmetry}- $USP(2N_c)$\cite{wittensmall}

{\it Hypermultiplets}-
(a)  $A$- antisymmetric rep.  $N_c(2N_c-1)$ 
which is irreducible $N_c(2N_c-1)$-1  $\oplus$ singlet.

(b)  $N_f=4$ $Q^i$,  ``quarks"  in the fundamental rep. $2N_c$
which emerge  in \Ft    from strings between 3B and the four  7B
at the orientifold point.

(iii){\it Global symmetry}-  $SO(8)$ global symmetry (local symmetry
in space-time)

(iv){\it Superpotential}-  $W=Q^iXQ^i + AXA + m_{ij}Q^iQ^j$

(v){\it Mass terms}-
$m_{ij}$ corresponds to the complex scalar 
$28$ atisymmetric rep. of the space-time $SO(8)$ laocal sym.  
No other scalars in space-time$\ra$ No mass term to $A$.
 
(vi){\it $\beta$ function}  $(2N_c+2)\ - (2N_c-2)\  -\ 4=0$

(vii){\it Expectation values}-
(a) scalars of the antisymmetric rep 
which correspond to the motion in  $x_4, x_5, x_6, x_7$. 
(b) scalars of vector multiplet
associated with Wilson loops of the gauge fields around $T^2$.    

(viii){\it Moduli space of Coulomb branch}-
$$ {(SU(2)^k)\over S_k}$$
parmutation symmetry $S_k$ part of the $USP(2N_c)$  Weyl group.  

\subsection{ Flows of the $USP(2N_c)$ theory}
 \ce{ (1) $USP(2N_c)\ \ra\ U(1)^{N_c}$ } 
$$<X> = diag(a_1,a_2,\cdots,a_k,-a_1,-a_2,\cdots,-a_k)$$
where $a_i^2$ corresponds in the \B picture to the
localtion of the $i^{th}$ 3B on ${T^2\over Z_2}$ from the
orientifold.


\begin{figure}[h]
\hbox to\hsize{\hss
\epsfxsize=6cm
\epsffile{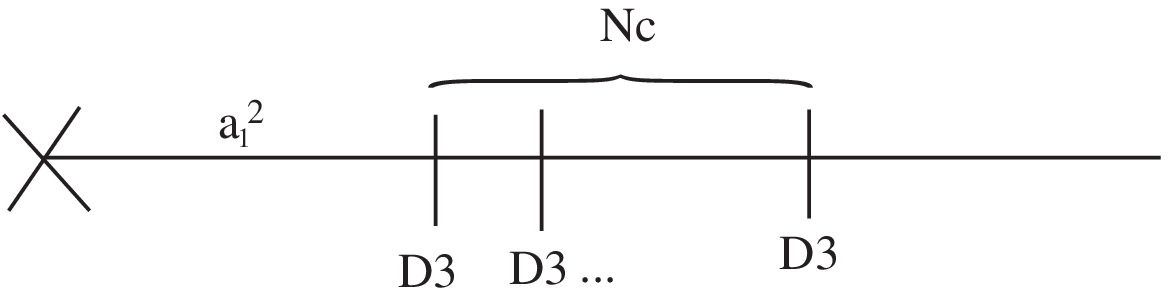} \hss}
\vspace{0cm}
\caption{} \label{pic1a}
\end{figure}

\ce{(2)  $USP(2N_c)\ \ra\ SU(2)^{N_c}$ } 
$$<A>=\pmatrix{0 & 0 & \cdots & 0 & b_1 & 0 & \cdots & 0 \cr
                      0 & 0 & \cdots & 0 & 0 & b_2 & \cdots & 0 \cr
         \vdots & \vdots & \ddots & \vdots & \vdots & \vdots & \vdots
& \vdots \cr
                      0 & 0 & \cdots & 0 & 0 & 0 & \cdots & b_k \cr
                      -b_1 & 0 & \cdots & 0 & 0 & 0 & \cdots & 0 \cr
                      0 & -b_2 & \cdots & 0 & 0 & 0 & \cdots & 0 \cr
     \vdots & \vdots & \vdots & \vdots & \vdots & \vdots & \ddots &
\vdots \cr
                     0 & 0 & \cdots & -b_k & 0 & 0 & \cdots & 0 \cr}.$$
$b_i^2$ corresponds in the \B picture to the
localtion of the $i^{th}$ 3B in $x_4,x_5,x_6$ and $x_7$.

In the limit of all $b_i$ being very large only 
singlet remanants from $A$  survives so it flows  to 
$N_c$ copies of Seiberg-Witten theory of $SU(2)$ with $N_f=4$. 

\ce{ (3) $USP(2N_c)\ \ra\ U({N_c}); N=4$ } 

The $N_c$ 3B are together but very far from the orientifold point
\cite{witbound}.

\begin{figure}[h]
\hbox to\hsize{\hss
\epsfxsize=6cm
\epsffile{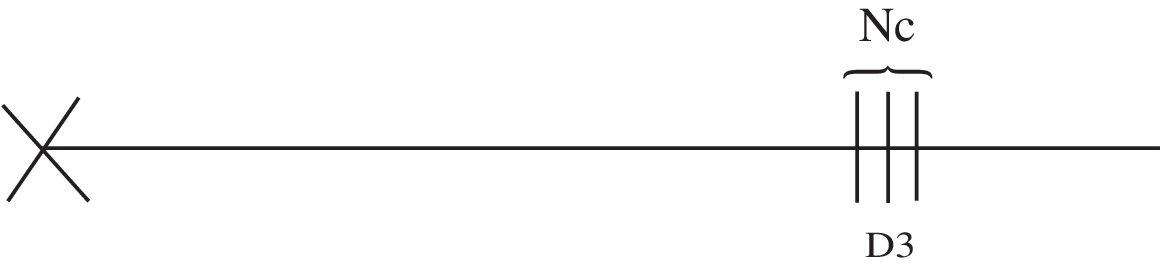} \hss}
\vspace{0cm}
\caption{} \label{pic1a}
\end{figure}

One can deduce from the probe theory interesting properties of the 
SW theories \cite{ASTY}
To summarize:

{\bf The  Coulomb phase of wv field theory of multiple $p$-branes 
is just  $N_c$ copies (up to global identifications)
of the Coulomb
phase  for a single $p$-brane.}

\section {Construction of $N=1$ superconformal
4D probe theory 
with electric-magnetic duality symmetry }

 Consider as in Sen's model\cite{sen}  
 compactifications of F-
theory on CY manifolds which have a constant 
 (weak coupling) value of $\tau$.

The simplest case  is
the elliptic fiber over a base $\bCP^1\times \bCP^1$\cite{mv} . 
\begin{figure}[h]
\hbox to\hsize{\hss
\epsfysize=3cm
\epsffile{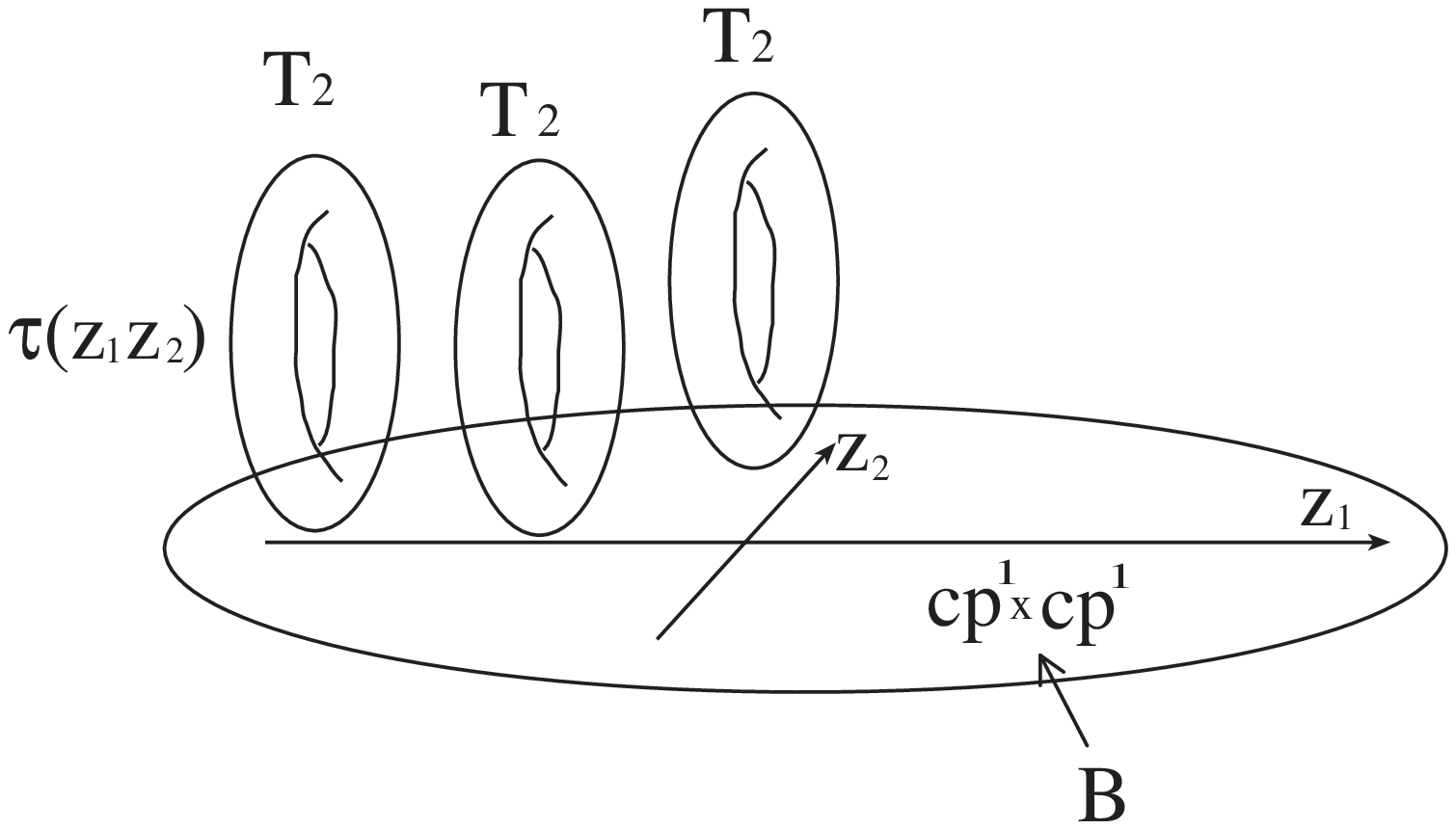} \hss}
\vspace{0cm}
\caption{} \label{pic1a}
\end{figure}
In analogy to (1) the toric fibers are described by 
$$\label{curve}y^2 = x^3 + f(z_1,z_2) x + g(z_1,z_2),$$
where $z_1$ and $z_2$ label the two $\bCP^1$s of the base.
 $$\Delta=4f^3+27g^2=0$$
At a degeneration of the fibers, which is a  
location of type IIB 7-brane in the
compact dimensions,
 $\tau$ has a non-trivial monodromy when going around it.

(i)  in F-theory 
$$\fbox{$\displaystyle \tau(z_1,z_2)=
constant\ra {f^3\over g^2}=constant$ }$$
A solution (though not the most general solution) which generalizes
\eqref{phi} is 
$$\fbox{$\displaystyle\label{choice}f(z_1,z_2) = \alpha \phi_1(z_1)^2 \phi_2(z_2)^2;
\ g(z_1,z_2) = \phi_1(z_1)^3 \phi_2(z_2)^3,$}$$

where $\phi_1$ and $\phi_2$ are general polynomials of degree four 
$$\fbox{$\displaystyle\label{phis}\phi_1(z_1) = \prod_{i=1}^4 (z_1 - z^{(i)}_1);
\ \phi_2(z_2) = \prod_{i=1}^4 (z_2 - z^{(i)}_2).$}$$

 At this point in moduli space $\tau=constant(\alpha)$ 
and there are  
$D_4$ singularities at $z_1=z_1^{(i)}$ and at $z_2=z_2^{(i)}$.

\subsection{ Space-time  theory
at the intersection of the two $D_4$ singularities }

In general the spacetime field theory at an intersection of
two $D_4$ singularities is not well understood\cite{bikmsv,bj}.
Let us  try to interpret this point in moduli space as an
orientifold of the type IIB theory.
Around each of the points $z_1=z^{(i)}_1$ (for
constant $z_2$) there is  an $SL(2,\bZ)$ monodromy of the form 
$$\label{monodromy}\pmatrix{-1 & 0 \cr 0 & -1 \cr}.$$
Locally, this monodromy  can be interpreted
 as an orientifold of the type IIB theory by $(-1)^{F_L}
\cdot \Omega$, 
 accompanied by four
7-branes to cancel the RR charge. 

Each point $z_1=z^{(i)}_1$
on the first $\bCP^1$ factor carries a
deficit angle of $\pi$, all four of them together
deforming the two  $\bCP^1$ to $\bT^2/\bZ_2$.
Thus, the theory looks like 
$$\fbox{$\displaystyle 
type\ II_B\ on\  {\bT^4\over \bZ_2\times \bZ_2'}$}$$
where
$$z_2={\cal I}_{67}\Omega (-1)^{F_L};\ \ 
z_2'={\cal I}_{89}\Omega (-1)^{F_L}. $$
Naively, this
corresponds to the orientifold of 
F-theory on ${\bT^6 \over \bZ_2\times \bZ_2'}$.

However, this orientifold
compactification
{\bf cannot be identified directly
 with the F-theory compactification\cite{curve}}

 Generally, in F-theory compactification  
$$B^{NS}_{\mu\nu}=B^{RR}_{\mu\nu}=0$$
In our case there is {\bf  discrete non-vanishing
$B_{\mu\nu} $fields } 
(related to {\it discrete torsion })
{\it How do we know?} 
At the
intersection of two $D_4$ singularities 
the intersection
point can be blown up  to get an additional 2-cycle. 
In the naive F-theory compactification
3-brane  can be wrapped over the 
vanishing 2-cycle, 
which should mean that  there are 
$\Rightarrow$ {\bf tensionless strings}
 living on the
intersection of the $D_4$ singularities.
 Our orientifold construction,
on the other hand, involves a well-defined weakly coupled conformal
field theory, whose low-energy spectrum is well-defined and can not
include such tensionless strings.
Discrete $\bZ_2$ symmetries force the value of both 2-form
tensor fields (integrated over the vanishing 2-cycle) to be either $0$
or $1/2$ (modulo $1$). 

A non-zero value for either (or both) of these fields breaks
the $SL(2,\bZ)$ symmetry to a discrete $\Gamma(2)$ subgroup.
 $$\Gamma(2); \{ s=\pmatrix{\ 0&1\cr -1&0\cr}, t^2= 
\pmatrix{\ 1&2\cr -0&1\cr} \}$$
There are two possible models:

(i) Integrated $B^{NS}={1\over 2}, B^{RR}=0$ $\ra${\bf 
Sen's $SU(4)$ model }\cite{sen2}

(ii) Integrated $B^{NS}=B^{RR}={1\over 2}$ $ \ra$ 
{\bf $SO(8)$ model}\cite{ASTY}
\subsection{ The space-time theory of the $SO(8)$ model} 
The superfield content which one  deduce from the various string
sectors are given by:
 
(1){\it  Untwisted closed string sectors} : 
 
 supergravity multiplet $+$   one tensor multiplet $+$ 
4 hypermultiplets. 

{\it Open string sector}: 

(i) 7-7 strings $\ra$ $SO(8)^4$  vector multiplets.

(ii) 7'-7' strings $\ra$ $SO(8)^4$  vector multiplets.

(iii) 7-7' no massless states.

(2) {\it  Twisted closed string sectors} :

A tensor multiplet at each
intersection of fixed points (i.e. at each fixed point of both
$\bZ_2$'s), 
accounting for a total of {\bf 16 additional tensor multiplets.} 
At each intersection of $D_4$ singularities
there is a possibility of blowing up a point to get an additional 2-cycle
After all these blow-ups we get

{\bf F-theory on a smooth $(h_{11}=51,h_{21}=3)$
Calabi-Yau manifold.}

which includes \cite{bz,dabpark,gm}:
$$n_T=h_{11}^B-1=18-1$$
$$n_{H^0}=h_{21}^M +1= 3+1=4$$
$$r(V)=h_{11}^M -h_{11}^B -1= 51-18-1=32$$
This set of 6D $N=1$ multiplets obey the anomaly cancellation condition
$$29 n_T + n_H -n_V=293$$ 
Note again, 
that the spacetime theory we found (using the orientifold
construction) is not the same as the theory we supposedly started
with, which was F-theory compactified on an elliptic fibration over
$\bCP^1 \times \bCP^1$.

\subsection{ The space-time theory of the $SU(4)$ model}\cite{sen2} 

{\it  Untwisted closed string sectors} : 
 
 supergravity multiplet $+$   one tensor multiplet $+$ 
4 hypermultiplets. 

{\it Open string sector}: 

(i) 7-7 strings $\ra$ $U(4)^4$  vector multiplets $+$ $2\times 6$
hypers for each vector.

(ii) 7'-7' strings $\ra$ $U(4)^4$  vector multiplets $+$ $2\times 6$
hypers for each vector.

(iii) 7-7'  $(4,4)$ hyper multiplets.

{\it  Twisted closed string sectors} :

16 hypers
$$ n_T=1,\ \  \ n_H=372\  n_v= 128$$
\subsection {3B probe wv field theory at the intersection}
The  probe theory is expected 
 (i) to have 
 only {\bf $N=1$ supersymmetry}.
( since the 7-branes intersect
transversely at this point and each breaks a different half of the
wv $N=4$ supersymmetry,)
(ii) to be invariant under conformal  and ($\Gamma(2)$)
electric-magnetic duality symmetry,
and (iii) to have two
deformations, corresponding to turning on $z_1$ or $z_2$, which
should cause it to flow to the $N=2$ $SU(2)$ 
gauge theory with 4 quark hypermultiplets.


\subsection{Fields from the 3-brane strings}
Let us begin by computing the fields
corresponding to strings which stretch out from a 3-brane and fold back to the
same brane. 

The 3-brane has 3 images under $\bZ_2 \times
\bZ_2$, so that every open string state is enhanced to a $4\times 4$
matrix (as in Gimon-Polchinski\cite{gp} (fig. 14 )
)

{\it  imposing  the orientifold restrictions.}
\begin{figure}[h]
\hbox to\hsize{\hss
\epsfysize=3cm
\epsffile{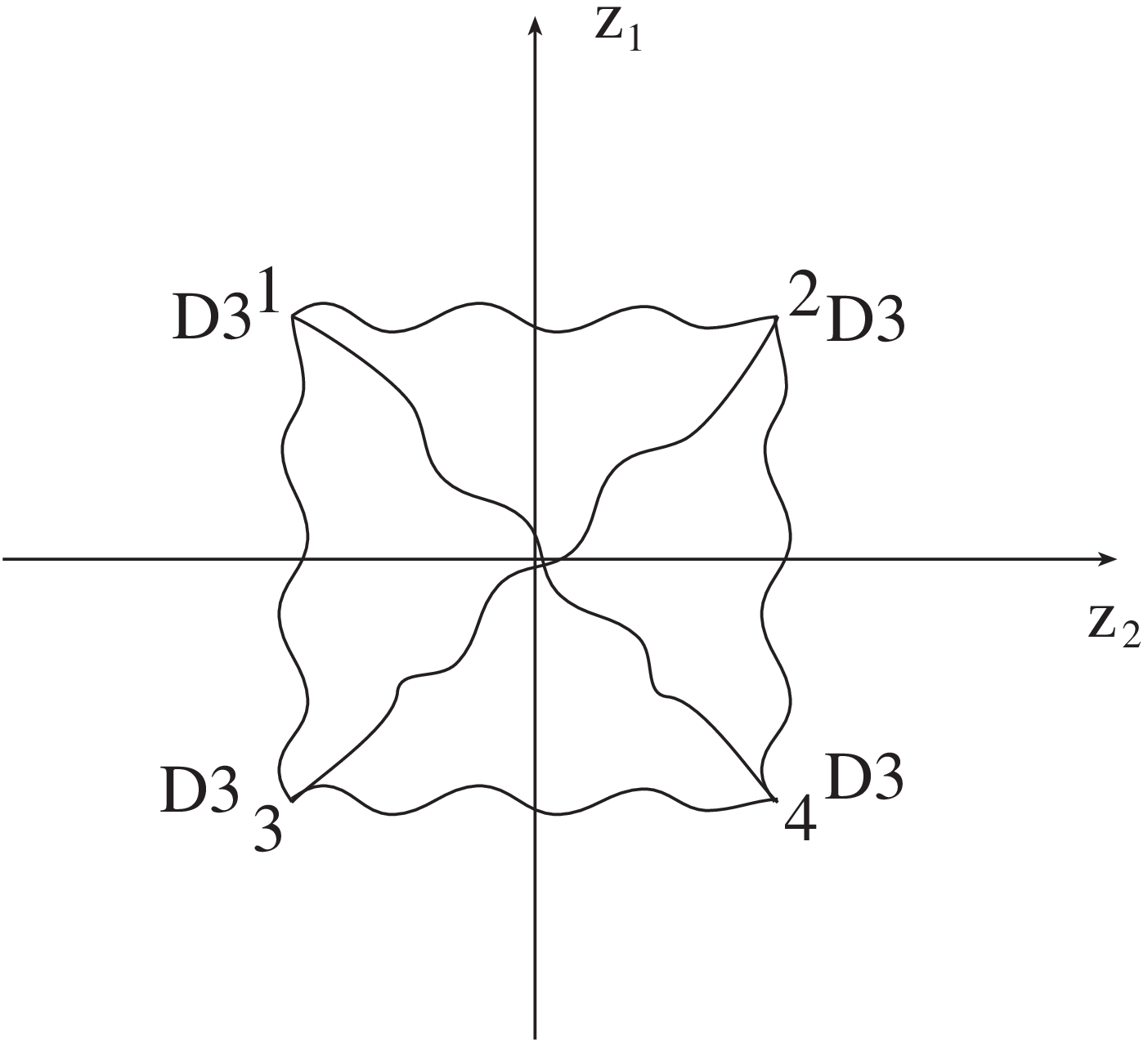} \hss}
\vspace{0cm}
\caption{} \label{pic1a}
\end{figure}

The $\gamma$ matrices\cite{gp} 
corresponding to each of the generators in the $\bZ_2 \times \bZ_2$
may be chosen to be
\ber
\gamma_{\Omega_1} = \pmatrix{ 0 & i & 0 & 0 \cr -i & 0 & 0 & 0 \cr
0 & 0 & 0 & i \cr 0 & 0 & -i & 0 \cr}  
\gamma_{\Omega_2} = \pmatrix{0 & 0 & i & 0 \cr 0 & 0 & 0 & i \cr -i &
0 & 0 & 0 \cr 0 & -i & 0 & 0 \cr}
\eer
and the orbifold matrix is then necessarily
\be
\label{gammatwo}
\gamma_T = \gamma_{\Omega_1} \gamma_{\Omega_2} = \pmatrix{0 & 0 & 0
& -1 \cr 0 & 0 & 1 & 0 \cr 0 & 1 & 0 & 0 \cr -1 & 0 & 0 & 0 \cr}.
\ee
These matrices are hermitian and unitary, and they correctly
reflect the two $\bZ_2$ actions on the compact coordinates $(z_1,z_2)$.

The wave-function matrices of states, $M_{ij}$, must then satisfy 
conditions of the type

\bear
\label{mishgamma}
M &= \pm \gamma_{\Omega_1} M^T
\gamma_{\Omega_1}^{-1} \cr
M &= \pm \gamma_{\Omega_2} M^T \gamma_{\Omega_2}^{-1} \cr
M &= \pm \gamma_T M \gamma_T^{-1} \cr
\eear

where the signs are determined by the transformation properties of the
relevant state.

For the gauge fields-  $-,-,+$, 

For the
chiral superfield $X_{6,7}$
- $-,+,-$, 

For  $X_{8,9}$ - $+,-,-$ 

 For $X_{4,5}$ - $+,+,+$.

Note that in flat space the 3-brane field theory involves an $N=4$
vector multiplet, containing an $N=1$ vector multiplet and three $N=1$
chiral multiplets corresponding to the coordinates $X_{4,5,6,7,8,9}$
of the 3-brane. In the presence of the orientifold, each of these
fields is enhanced into a $4\times 4$ matrix with different
constraints.

\medskip
\centerline{(i) \bf Vector superfields}

 The relations imposed by \eqref{mishgamma}\ on the
components $M_{ij}$  of gauge fields are
\bear
 M_{11}&= -M_{22} = -M_{33} = M_{44} \cr
 M_{14}&= M_{41} = M_{23} = M_{32} \cr
 M_{12}&= -M_{43} \quad M_{13} = -M_{42} \cr
M_{21}&= -M_{34} \quad M_{31} = -M_{24}. \cr
\eear

A basis of 6 matrices that obey these relations is
\ber
Z_1 = \pmatrix{ 1 & 0 & 0 & 0 \cr 0 & -1 & 0 &
0 \cr 0 & 0 & -1 & 0 \cr 0 & 0 & 0 & 1 \cr}   &
Z_2 = \pmatrix{0 & 0 & 0 & -1 \cr 0 & 0 & -1 & 0 \cr 0 &
-1 & 0 & 0 \cr -1 & 0 & 0 & 0 \cr} \cr
W_1 = \pmatrix{ 0 & 1 & 0 & 0 \cr 1 & 0 & 0 & 0 \cr
0 & 0 & 0 & -1 \cr 0 & 0 & -1 & 0 \cr}   &
W_2 = \pmatrix{0 & 0 & 1 & 0 \cr 0 & 0 & 0 & -1 \cr 1 &
0 & 0 & 0 \cr 0 & -1 & 0 & 0 \cr} \cr
W_3 = \pmatrix{ 0 & 0 & i & 0 \cr 0 & 0 & 0 & i \cr
-i & 0 & 0 & 0 \cr 0 & -i & 0 & 0 \cr}   &
W_4 = \pmatrix{0 & i & 0 & 0 \cr -i & 0 & 0 & 0 \cr 0 &
0 & 0 & i \cr 0 & 0 & -i & 0 \cr}.
\eer

It is now straightforward to check that the matrices

\ber
W^+_2=&{1\over 4}(W_1 + W_2);\ W^-_1={1\over 4}(W_1 -
W_2);\cr
 \ W^+_1=&{1\over 4}(W_3 + W_4); \ W^-_2={1\over 4}(W_3 -
W_4) \cr Z^{\pm}=&{1\over 4}(Z_1\pm Z_2)\cr
\eer
obey the following commutation relations
\ber
\label{mishcr} [Z^+, W^+_2]=&-i W^+_1;\ \  [Z^+, W^+_1]=i
W^+_2;\cr
 \  [W^+_2, W^+_1]=&-i Z^+\cr [Z^-, W^-_1]=&i W^-_2;\cr \  [Z^-,
W^-_2]=&-i W^-_1;\ \ [W^-_1, W^-_2]=i Z^-\cr [Z^+, W^-_1]=&0;\ \
[Z^+, W^-_2]=0;\cr \ [Z^-, W^+_2]=&0;\ \  [Z^-, W^+_1]=0;\ \
[Z^+,Z^-]=0\cr [W^+_2, W^-_1]=&0;\ \  [W^+_2, W^-_2]=0;\cr \ [W^+_1,
W^-_1]=&0;\ \  [W^+_1, W^-_2]=0. \cr
\eer
Thus, we see that the gauge fields span an

\ce{\bf $SU(2) \times SU(2)$ algebra.}

\begin{figure}[h]
\hbox to\hsize{\hss
\epsfysize=3cm
\epsffile{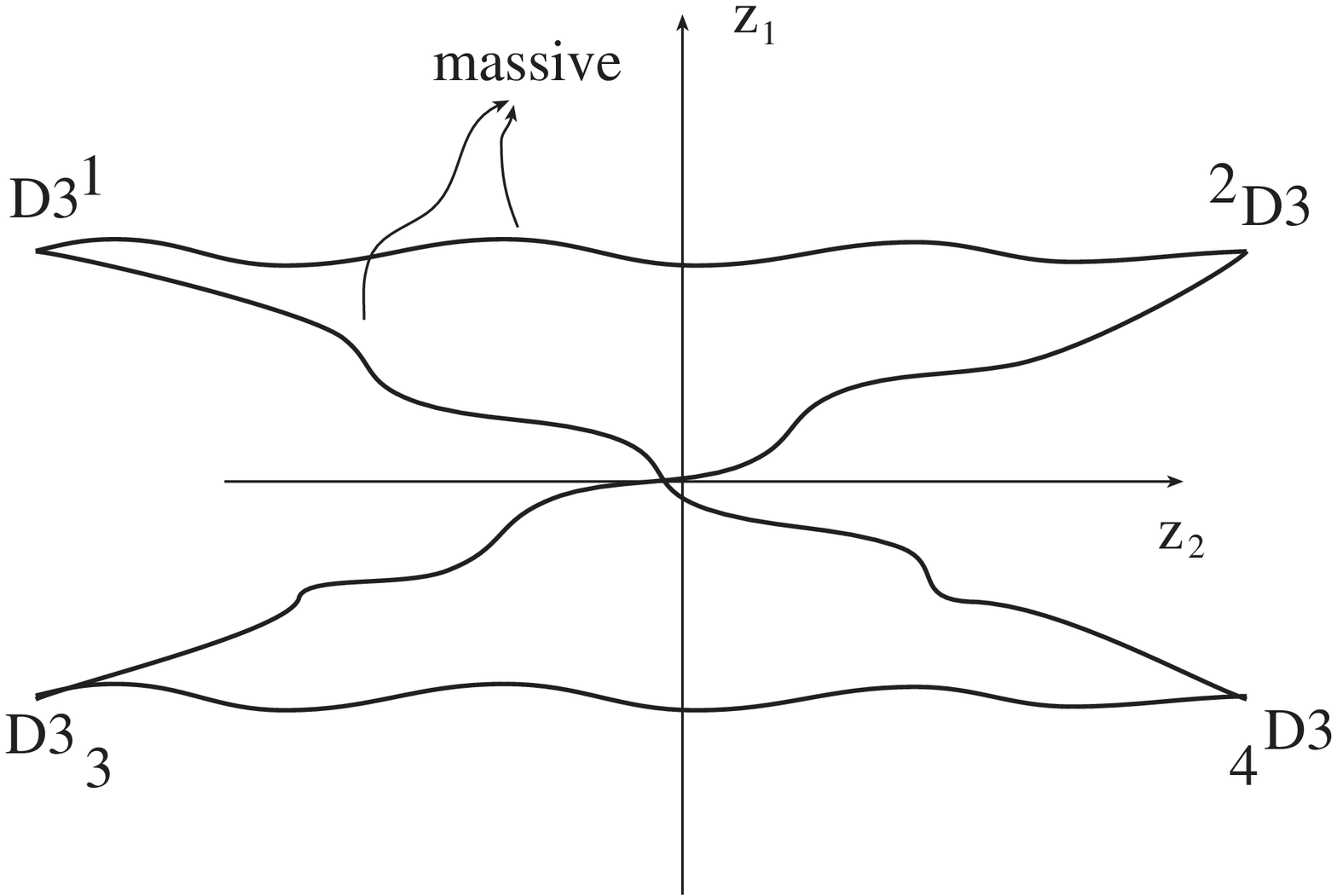} \hss}
\vspace{0cm}
\caption{} \label{pic1a}
\end{figure}

When we take $z_1$ to infinity all 1-2,1-4,2-3 and 3-4 strings become
infinitely massive, (fig. 15)  
 so we can drop those wave-function matrices
with non-zero entries in those positions. 
This leaves
$Z_1,\,W_2$ and $W_3$, which generate an {\bf  $SU(2)$ subalgebra.}

Likewise, if we take $z_2$ to infinity.
This  picture is in accord with the naive expectation following Sen
of having just a single $SU(2)$ near $z_1=0$ or $z_2=0$.

\medskip
\centerline{(ii) \bf $X_{67}$ chiral multiplet}

Next consider the  implications of \cite{mishgamma}\ on the scalar
fields   $X_{67}$.  The constraints on the
matrix components are now
\ber
\label{mishconsb} M_{11}&= -M_{22} = M_{33} = -M_{44} \cr
 M_{14}&= -M_{41} = -M_{23} = M_{32} \cr
 M_{12}&= +M_{43} \quad  M_{13} = M_{42}=0 \cr
M_{21}&= +M_{34} \quad  M_{31} = M_{24}=0 \cr
\eer
A basis of hermitian matrices that obey the relations is the following
\ber
\label{xsixmats}A_1 = \pmatrix{ 1 & 0 & 0 & 0 \cr 0 & -1 & 0 &
0 \cr 0 & 0 & 1 & 0 \cr 0 & 0 & 0 & -1 \cr}   &
A_2 = \pmatrix{0 & 0 & 0 & i \cr 0 & 0 & -i & 0 \cr 0 &
i & 0 & 0 \cr -i & 0 & 0 & 0 \cr} \cr
A_3 = \pmatrix{ 0 & 1 & 0 & 0 \cr 1 & 0 & 0 & 0 \cr
0 & 0 & 0 & 1 \cr 0 & 0 & 1 & 0 \cr}   &
A_4 = \pmatrix{0 & i & 0 & 0 \cr -i & 0 & 0 & 0 \cr 0 &
0 & 0 & -i \cr 0 & 0 & i & 0 \cr}. \cr
\eer
Note that now if we take $z_1$ to infinity (naively) we are left only
with $A_1$, while if we take $z_2$ to infinity (naively) we remain
with $A_1,A_3$ and $A_4$ which are in the adjoint representation of
the remaining $SU(2)$, and thus we go over to the  picture of Sen,
as expected.

Define now the  matrices

$A_{-+},A_{++},A_{--}$ and $A_{+-}$ 

by the equations

$A_{++}=A_3-iA_4$;\  $A_{-+}=A_1+iA_2$;\ $A_{+-}=A_1-iA_2$ and
$A_{--}=A_3+iA_4$.  

It can easily be checked that

$A= (A_{++},A_{-+},A_{+-},A_{--})$ are in the

{\bf $(2,2)$ representation of $SU(2) \times SU(2)$} 
 For instance, we
have
$$ [Z^+,A_{++}] = \ha A_{++} \ \ [Z^+,A_{-+}] = -\ha A_{-+} $$
$$ [W^+_2,A_{++}] = -\ha A_{-+} \ \ [W^+_2,A_{-+}] = -\ha A_{++} $$
$$ [W^+_1,A_{++}] = {i\over 2} A_{-+} \ \ [W^+_1,A_{-+}] = -{i\over 2}
A_{++} $$
and so on.

\medskip
\centerline{(iii) \bf $X_{89}$ chiral multiplet}

For the relations on   the $X_{89}$ chiral superfield
we would find similar (but not identical)
results to those of  $X_{67}$
(with the second and third rows and columns of all
matrices interchanged.)

Thus, this chiral superfield, that we denote by
$$B= (B_{++},B_{-+},B_{+-},B_{--})$$,
is also
in the 
{\bf$(2,2)$ representation of $SU(2) \times SU(2)$}

As a consistency check, we verify
that a VEV for $X_{67}$, for instance, indeed breaks the gauge
symmetry to a diagonal $SU(2)$. For instance, taking $A_1$ to have
a VEV would leave exactly the matrices $Z_1,W_2$ and $W_3$ (given
above), which commute with it, as expected. The $A$'s would also all
become massive except $A_1$ (since they do not commute with it),
again as expected (since after moving along the flat direction we
should have just a single scalar).

\medskip
\centerline{(iv) \bf $X_{45}$ chiral multiplet}

The relations for $X_{45}$ are
\ber
\label{mishconsb} M_{11}&= M_{22} = M_{33} = M_{44} \cr
 M_{14}&= M_{41} = -M_{23} = -M_{32}, \cr
\eer
and their solutions are spanned by the two singlets $S_1$ and $S_2$
\ber
\label{xfourmats}S_1 = \pmatrix{ 1 & 0 & 0 & 0 \cr 0 & 1 & 0 & 0 \cr
0 & 0 & 1 & 0 \cr 0 & 0 & 0 & 1 \cr}  \quad
S_2 = \pmatrix{0 & 0 & 0 & 1 \cr 0 & 0 & -1 & 0 \cr 0 &
-1 & 0 & 0 \cr 1 & 0 & 0 & 0 \cr}
\eer
\subsection{Fields form the strings between the 3-brane and 7-branes }

The
 $\gamma_{\Omega}$ matrices for the
7-branes  are
all proportional to the identity matrix 
The constraint on the
7-brane gauge fields is then just 
$$M = - M^T,$$ 
giving an anti-symmetric
matrix corresponding to an $SO(8)$ space-time
gauge symmetry, since there are 8 7-branes
when including the $Z_2$ partners.

Next, we should use these matrices to analyze the wave function
of the 3-7 strings (as in \cite{gp}). 
The 7-brane side is
trivial, so the orbifold/orientifold projections
just mix the various 3-branes
according to the $\gamma_{\Omega}$ matrices
For instance  taking
the $3_1-7$ state (where $3_1$ is the first 3-brane) via
$\gamma_{\Omega_1}$ to $i$ times
the $7-3_2$ state (with opposite orientation !),
via $\gamma_{\Omega_2}$ to $i$ times the $7-3_3$
state, and via $\gamma_{T}$ to minus the $3_4-7$ state.
A basis for the states going to a
specific 7-brane can thus be chosen to be
$$\label{quarks}D^+_u=(1,0,0,-1); \qquad D^+_d (0,1,1,0).$$
Using the matrices we found above for the gauge fields,
it is easy to check that these are doublets under $SU(2)_+$, and
singlets with respect to
$SU(2)_-$. 
The corresponding chiral superfields are thus in
 {\bf$(2,1)$ representation of
$SU(2)_+ \times SU(2)_-$.}
For the other group of 7-branes we can then do
the same thing, but with minus the identity matrix for some of the
relevant 7-7 $\gamma_{\Omega}$ matrices (say for
$\gamma_{\Omega_2}$ and $\gamma_T$). 
Then, the basis
comes out to be
$$D^-_u=(1,0,0,1); \qquad D^-_d (0,1,-1,0). $$
which is in the 
{\bf$(1,2)$
representation of  $SU(2)_+ \times SU(2)_-$.}

The summary of the field content is  

{\bf  $A_\mu$--- $SU(2) \times SU(2)$ gauge fields.
\hfill 

$X_{8,9}\equiv B$--- chiral field in $(2,2)$\ \ \  $A^2=z_2$  
\hfill 

$X_{6,7}\equiv A$--- chiral field in $(2,2)$\ \ \  $A^2=z_1$ 
\hfill  

$X_{4,5}\equiv (S_1,S_2)$---  singlet chiral fields 
\hfill 

$Q^i$ ($i=1,\cdots,8$)---- chiral quarks in $(2,1)$
\hfill 

$q^i$ ($i=1,\cdots,8$)---- chiral quarks in $(1,2)$}

{\it Note about this particle spectrum.}

(i) It is manifestly not $N=2$ supersymmetric (since
there is no chiral multiplet in the adjoint representation)

(ii) $<A>\neq 0$ ($<B>\neq 0$) breaks
$SU(2)\times SU(2)\ra SU(2)$-

Three components of $A$ ($B$) are
swallowed by the Higgs mechanism, and we remain with an adjoint chiral
multiplet (coming from $B$ ($A$)) and additional singlets, as
expected. 

(iii) A natural interpretation of $S_1$ $S_2$
is that at the intersection point
the 3-brane can split into two half-3-branes which can move
independently. 

(iv) {\bf Vanishing one loop $\beta$ function }

(since there are 12 ($N_f=6$) doublets of each $SU(2)$).
\subsection{ The Superpotential}

Let start with some  general points:

(i) In
principle, it should be  possible to compute this superpotential
from the string theory analysis, but we have not performed this
computation.

(ii) Since we have only $N=1$ supersymmetry, the
superpotential in general receives non-perturbative quantum corrections.

(iii) The superpotential has to obey the following main constraints:

1) The ($SO(8)$ theory should have (at least) an $SO(8) \times SO(8)$
global symmetry.
The ($SU(4)$) theory should have (at least) an $SU(4) \times SU(4)$
global symmetry.

2) The theory should flow to the $N=2$, $SU(2)$, $N_f=4$ theory upon
giving a VEV to $A$ or to $B$.

3) The theory should have a fixed line passing through weak coupling
\cite{ls,aks}.

4) The quarks $q^i$ ($Q^i$) have to become
massive when we give a VEV to $A$ ($B$)


{\it Construction of the superpotential}:

(i) {\bf  Superpotential for the quarks}
The vanishing of the
$\beta$ function associated with the gauge coupling $\beta_g$ 
and with the superpotential couplings $\beta_{h_n}$  requires that 
\bear
A_g=&3C_2(G) -\sum_i T(r_i) + \sum_i T(r_i)\gamma_i=0\\
A_{h_n}=& (n-3) +{1\over 2}\sum_k\gamma_k=0\\
\eear
where in the first equation the sum is over all the fields in 
the theory, and in the second only over the fields in the
particular term of the superpotential. 

A fixed line passing through weak coupling demands that these
equations are not all independent\cite{ls}. 
It is easy to check tht $n=3$ and $\gamma_k=0$ is a solution for 
this condition. In fact in \cite{aks} it was shown from string
argumetns that indeed all the anomalous dimensions vanish.

Combining this solution with  the requirements 1), 2) and 4) 
one ends up with the following superpotentials

For the $SU(4)$ model\cite{sen2}

\be 
W=QA\tilde Q +  qB\tilde q  
\ee
and for the case of $SO(8)$\cite{ASTY}

\be
W= {QABQ\over{\sqrt{A^2}}}+ {qABq\over{\sqrt{B^2}}} 
\ee
Inspite of the fact that in both cases these superpotentials obey
all the requirements, it is clear that only for the   $SU(4)$ model
it is an adequate solution. In the second case one encounters
singularities in the limits of $A\ra 0$ and $B\ra 0$ which cannot
be smoothed .

(ii) {\bf Construction of the superpotential for $S_1,S_2)$}
\noindent

(a) $S_1$-  the
location of the 3-brane in $X_{4,5}$ coordinates should be decoupled.

(b) $S_2$ (corresponding to splitting the 3-brane at the
orientifold point) should be massless only when $A$ and $B$ are both
zero.

(c) $A$ and $B$ become massive once
$S_2$ is turned on, as expected, since when the 3-brane has split we
cannot move it away from the orientifold point.

(d) In the ``Coulomb branch'' $<A^2>$ and $<B^2>$ should
correspond to flat directions but not $AB$.

(e) $<A>$-three of its components are swallowed by the Higgs
mechanism, and the other remains massless and parametrizes the flat
direction corresponding to the flow (it is the $N=2$ partner of
$S_1$). Three of the components of $B$
remain massless and become an adjoint field $X$ of the remaining $SU(2)$,
but the remaining component, as well as $S_2$, should become massive 
(since we have no corresponding fields in the $N=2$ theory)
The final result for this part of the action is
$$ W = S_2 A B$$

{\bf Acknowledgements}
We would like to thank A. Brandhuber, V. Kaplunovsky and especially
A. Sen for useful conversations.

\def\np#1#2#3{Nucl. Phys. {\bf B#1} (#2) #3}
\def\pl#1#2#3{Phys. Lett. {\bf #1B} (#2) #3}
\def\prl#1#2#3{Phys. Rev. Lett. {\bf #1} (#2) #3}
\def\cmp#1#2#3{Comm. Math. Phys. {\bf #1} (#2) #3}
\def\physrev#1#2#3{Phys. Rev. {\bf D#1} (#2) #3}
\def\jgp#1#2#3{J. Geom. Phys. {\bf #1} (#2) #3}


\end{document}